\documentclass[11pt,a4paper,english]{article}
\usepackage[round,sort]{natbib}
\usepackage[latin1]{inputenc}
\usepackage[T1]{fontenc} 
\usepackage[english]{babel}
\usepackage{amsmath,amsfonts,dsfont,amsthm,amssymb,amsbsy,graphicx,graphics,pdfpages, geometry}
\usepackage{subfigure}
\usepackage{booktabs}
\newcommand\T{\rule{0pt}{2.6ex}}       
\newcommand\B{\rule[-1.2ex]{0pt}{0pt}} 

\theoremstyle{definition}

\newtheorem{scheme}{Scheme}

\title{\textbf{A similarity-based implementation of the Schaake shuffle}}
\author{Roman Schefzik\vspace{0.2 cm}\\ \textit{Heidelberg Institute for Theoretical Studies}\\ \textit{Schloss-Wolfsbrunnenweg 35, 69118 Heidelberg, Germany}\vspace{0.2 cm}\\ \texttt{roman.schefzik@h-its.org}}
\date{\today}

\begin{document}

\maketitle

\begin{abstract}
Contemporary weather forecasts are typically based on ensemble prediction systems, which consist of multiple runs of numerical weather prediction models that vary with respect to in the initial conditions and/or the the parameterization of the atmosphere. Ensemble forecasts are frequently biased and show dispersion errors and thus need to be statistically postprocessed. However, current postprocessing approaches are often univariate and apply to a single weather quantity at a single location and for a single prediction horizon only, thereby failing to account for potentially crucial dependence structures. Non-parametric multivariate postprocessing methods based on empirical copulas, such as ensemble copula coupling or the Schaake shuffle, can address this shortcoming. A specific implementation of the Schaake shuffle, called the SimSchaake approach, is introduced. The SimSchaake method aggregates univariately postprocessed ensemble forecasts using dependence patterns from past observations. Specifically, the observations are taken from historical dates at which the ensemble forecasts resembled the current ensemble prediction with respect to a specific similarity criterion. The SimSchaake ensemble outperforms all reference ensembles in an application to ensemble forecasts for surface temperature from the European Centre for Medium-Range Weather Forecasts.
\end{abstract}

\textit{Keywords:} empirical copula, ensemble copula coupling, probabilistic weather forecasting, Schaake shuffle, similarity criterion, statistical ensemble postprocessing

%








\section{Introduction}\label{intro}

Contemporary weather forecasts are typically constructed from ensemble prediction systems, which have run operationally since the early 1990s. An ensemble consists of multiple runs of numerical weather prediction models which vary with respect to the initial conditions and/or the parameterization of the atmosphere. Consequently, ensembles take account of the two major sources of uncertainty \citep{Palmer2002,GneitingRaftery2005,LeutbecherPalmer2008}. Ensemble forecasts are frequently biased and show dispersion errors \citep{HamillColucci1997}. Thus, they require statistical postprocessing to realize their full capability. During the last decade, several ensemble postprocessing methods have been developed. Examples are (variants of) the ensemble model output statistics (EMOS) \citep[among others]{Gneiting&2005} approach, which is also known as non-homogeneous regression, or Bayesian model averaging (BMA) \citep[among others]{Raftery&2005}. However, EMOS and BMA, as well as other postprocessing methods, only apply to a single weather variable at a single location and for a single prediction horizon. Thus, they fail to account for spatial, temporal or inter-variable dependence structures, which are crucial in many applications such as flood warning \citep{Schaake&2010}, winter road maintenance \citep{Berrocal&2010} or the handling of renewable energy sources \citep{Pinson2013}. In recent years, there has been a keen interest in the development of multivariate postprocessing methods being able to address this shortcoming, and a lot of effort has been invested to that end. For instance, variants and modifications of EMOS and BMA, respectively, that can handle spatial \citep{Berrocal&2007,Berrocal&2008,Feldmann&2014} or inter-variable \citep{Schuhen&2012,Moeller&2013,Sloughter&2013,BaranMoeller2014} dependencies are available. Furthermore, there are methods to capture temporal dependencies of consecutive lead times in postprocessed predictive distributions \citep{Pinson&2009,SchoelzelHense2011}. All these multivariate postprocessing methods are parametric and work well in rather low dimensions and in settings in which the involved correlation matrix can be taken to be highly structured. However, they model one type of dependence (spatial, inter-variable or temporal) only and appear to be inadequate when considering high-dimensional situations, in which no particular structure can be exploited. These issues can be addressed using non-parametric techniques based on the use of empirical copulas \citep{Schefzik2015b}. Following \citet{Wilks2014}, an empirical copula \citep{Deheuvels1979,Rueschendorf2009} can be interpreted as a dependence template induced by a specific discrete multivariate data set. It can be employed to transfer a particular dependence pattern to samples which are drawn independently from a collection of univariate marginal distributions \citep{Wilks2014}. For example, in the ensemble copula coupling (ECC) approach of \citet{Schefzik&2013}, such an empirical copula is derived from the unprocessed ensemble forecast and then applied to samples from univariate postprocessed predictive distributions, which can be gained via the standard EMOS or BMA approaches. This is equivalent to ordering these samples according to the rank dependence structure of the raw ensemble, thereby capturing the spatial, temporal and inter-variable flow dependence \citep{Schefzik&2013}. Proceeding in a similar manner, the Schaake shuffle \citep{Clark&2004} employs an ordering based on past observations from a historical data archive. Consequently, the corresponding empirical copula in the Schaake shuffle is induced by an observational database rather than by an ensemble forecast. However, the standard Schaake shuffle fails to condition the multivariate dependence pattern on current or predicted atmospheric states. To address this shortcoming, \citet[page 260]{Clark&2004} proposed to develop an extension thereof, driven by the idea
\begin{quote}
``to preferentially select dates from the historical record that resemble forecasted atmospheric conditions and use the spatial correlation structure from this subset of dates to reconstruct the spatial variability for a specific forecast.''
\end{quote}
Inspired by this suggestion, a specific implementation of the Schaake shuffle, referred to as the SimSchaake approach, is introduced in this paper. Essentially, the SimSchaake method proceeds like the Schaake shuffle, but the observations determining the dependence structure are taken from historical dates at which the ensemble forecasts resembled the current ensemble prediction with respect to a specific similarity criterion.
\\
The remainder of the paper is organized as follows. In Section \ref{methods}, we first discuss the general setting of empirical copula-based ensemble postprocessing and review ECC and the Schaake shuffle as reference examples. Then, we develop the SimSchaake approach. In Section \ref{case.study}, this new method is evaluated and compared to the reference methods in a case study. The paper closes with a discussion in Section \ref{discussion}.

\section{Empirical copula-based ensemble postprocessing methods}\label{methods}

Copulas are valuable and established tools for the modeling of stochastic dependence \citep{Nelsen2006,Joe2014}. They have been successfully employed in numerous application areas. A copula is an $L$-variate cumulative distribution function (CDF) with standard uniform univariate marginal CDFs, where $L \in \mathbb{N}$, $L \geq 2$. As is manifested in the famous Sklar's theorem \citep{Sklar1959}, a copula $C$ links a multivariate CDF $H$ to its univariate marginal CDFs $F_1,\ldots,F_L$ via the decomposition
\begin{equation*}
H(u_1,\ldots,u_L)=C(F_1(u_1),\ldots,F_L(u_L)) 
\end{equation*}
for $u_1,\ldots,u_L \in \mathbb{R}$. In a multivariate postprocessing setting, the sought multivariate CDF $H$ can thus be constructed by specifying both the univariate marginal CDFs $F_1,\ldots,F_L$ and the copula $C$ modeling the dependence. The CDFs $F_1,\ldots,F_L$ can be obtained by common univariate postprocessing for each location, weather variable and look-ahead time individually, for instance performed via EMOS or BMA. For the choice of $C$, a prominent example is the Gaussian copula, which has been applied to a wide range of problems in climatology, meteorology and hydrology \citep{Schuhen&2012,Pinson2012,Moeller&2013,GenestFavre2007,SchoelzelFriederichs2008}.\\
In this paper, we focus on the case in which $F_1,\ldots,F_L$ are the empirical CDFs given by samples from univariate postprocessed CDFs and $C$ is taken to be an empirical copula \citep{Deheuvels1979,Rueschendorf2009}, $E_N$. According to \citet{Wilks2014}, an empirical copula can be considered a multivariate dependence template derived from a specific discrete data set. To describe this formally, let ${\mathbf{z}}:=\{(z_{1}^{1},\ldots,z_{N}^{1}),\ldots,(z_{1}^{L},\ldots,z_{N}^{L})\}$ be a data set consisting of $L$ tuples of size $N$ with entries in $\mathbb{R}$. Moreover, let $\operatorname{rank}(z_n^\ell)$ denote the rank of $z_n^\ell$ in $\{z_1^\ell,\ldots,z_N^\ell\}$ for $n \in \{1,\ldots,N\}$ and $\ell \in \{1,\ldots,L\}$, assuming for simplicity that there are no ties. Then, the empirical copula $E_{N}$ induced by the data set ${\mathbf{z}}$ is given by
\begin{eqnarray}\label{empcop}
E_{N}\left(\frac{i_{1}}{N},\ldots,\frac{i_{L}}{N}\right) &:=&\frac{1}{N} \sum\limits_{n=1}^{N}\mathds{1}_{\{\operatorname{rank}(z_{n}^{1}) \leq i_1,\ldots,\operatorname{rank}(z_{n}^{L})\leq i_L\}}\\ &=& \frac{1}{N}  
\sum\limits_{n=1}^{N} \prod\limits_{\ell=1}^{L}  
\mathds{1}_{\left\{\operatorname{rank}(z_{n}^{\ell}) \leq i_{\ell}\right\}}\nonumber
\end{eqnarray}
for integers $0 \leq i_1,\ldots,i_L \leq N$, with $\mathds{1}_A$ denoting the indicator function whose value is 1 if the event $A$ occurs, and zero otherwise.
\\
In this section, we review ensemble copula coupling (ECC) \citep{Schefzik&2013} and the Schaake shuffle \citep{Clark&2004} as reference methods within the general frame of empirical copula-based multivariate ensemble postprocessing \citep{Schefzik2015b}. In addition, we develop the SimSchaake method as a specific implementation of the Schaake shuffle. While ECC and the Schaake shuffle have been used as a benchmark in several papers \citep{Schaake&2007,ScheuererHamill2014,Voisin&2011,VracFriederichs2014,Wilks2014}, the SimSchaake method is new.\\
To allow for a formal description of the approaches, let us first set some notation which will be used throughout the whole section. Let  $i \in \{1,\ldots,I\}$ be a weather variable, $j \in \{1,\ldots,J\}$ a location and $k \in \{1,\ldots,K\}$ a look-ahead time. For simplicity, let $\ell:=(i,j,k)$ and $\ell^{\ast}:=(i,j)$ denote the corresponding multi-indices, and let $L:=I \times J \times K$ and $L^{\ast}:=I \times J$, respectively. Moreover, let $M$ denote the number of raw ensemble members, and $N$ the desired number of members the postprocessed ensemble shall consist of.
\\
In general, multivariate empirical copula-based ensemble methods to postprocess a raw ensemble forecast ${\mathbf{x}}:=\{(x_{1}^{1},\ldots,x_{M}^{1}),\ldots,(x_{1}^{L},\ldots,x_{M}^{L})\}$ initialized at a specific date $t_0$ proceed according to the following scheme.
\begin{scheme}\label{scheme.empcop}(Empirical copula-based multivariate ensemble postprocessing)
\begin{enumerate}
\item{To implement dependence structures, derive an empirical copula $E_N$ from a suitable data set ${\mathbf{z}}:=\{(z_{1}^{1},\ldots,z_{N}^{1}),\ldots,(z_{1}^{L},\ldots,z_{N}^{L})\}$ via \eqref{empcop}.\\ Equivalently, derive the univariate order statistics $z_{(1)}^{\ell} \leq \ldots \leq z_{(N)}^{\ell}$ for $\ell \in \{1,\ldots,L\}$, inducing the permutation $\pi_{\ell}(n):=\operatorname{rank}(z_n^{\ell})$ for $n \in \{1,\ldots,N\}$. If there are ties, let them be resolved at random.}
\item{For each margin $\ell$, perform univariate postprocessing of the raw ensemble forecast $x_1^\ell,\ldots,x_M^\ell$ based on past forecasts and observations as training data, and obtain a postprocessed predictive CDF $F_\ell$ in each case.}
\item{Draw a sample $\tilde{x}_{1}^{\ell},\ldots,\tilde{x}_{N}^{\ell}$ from each marginal CDF $F_\ell$. This can be for instance and most conveniently done by taking equidistant quantiles of $F_\ell$ of the form
\begin{equation}\label{sampling.q}
\tilde{x}_{1}^{\ell}:=F_\ell^{-1}\left(\frac{1}{N+1}\right),\ldots,\tilde{x}_{N}^{\ell}:=F_\ell^{-1}\left(\frac{N}{N+1}\right).
\end{equation}
}
\item{Apply the empirical copula $E_N$ to the samples from step 3.\\ Equivalently, arrange these samples with respect to the rank dependence structure of the chosen data set ${\mathbf{z}}$ in step 1. Using the permutation $\pi_\ell$, the final postprocessed ensemble $\hat{x}_{1}^{\ell},\ldots,\hat{x}_{N}^{\ell}$ for each margin $\ell$ is thus given by
\begin{equation*}
\hat{x}_{1}^{\ell}:=\tilde{x}_{(\pi_\ell(1))}^{\ell},\ldots,\hat{x}_{N}^{\ell}:=\tilde{x}_{(\pi_\ell(N))}^{\ell}.
\end{equation*}
}
\end{enumerate}
\end{scheme}

\subsection{Reference methods}\label{ref.methods}

Now we review ECC \citep{Schefzik&2013} and the Schaake shuffle \citep{Clark&2004} as empirical copula-based postprocessing techniques within the scheme described above.

\subsubsection{Ensemble copula coupling (ECC)}\label{ecc}

In the ECC approach \citep{Schefzik&2013}, the data set specifying the dependence structure is given by the raw ensemble forecast, that is, we have $\mathbf{z}=\mathbf{x}=\{(x_{1}^{1},\ldots,x_{M}^{1}),\ldots,(x_{1}^{L},\ldots,x_{M}^{L})\}$ in Scheme \ref{scheme.empcop}. Consequently, the sample size in step 3 and hence the size of the final postprocessed ensemble in step 4 of Scheme \ref{scheme.empcop} is restricted to equal that of the unprocessed ensemble, that is, $N=M$.\\
The ECC procedure operates under a perfect model assumption, implictly assuming that the ensemble is capable to represent actual spatial, inter-variable and temporal dependence structures adequately. This may or may not be expected, but surely does not hold each and every day. Moreover, ECC only applies to ensembles whose members can be considered exchangeable, that is, statistically indistinguishable.\\
\citet{Schefzik&2013} and \citet{Schefzik2015} show that ECC provides an overarching frame for seemingly unrelated approaches scattered in the literature such as those of \citet{Pinson2012}, \citet{RoulinVannitsem2012}, \citet{Flowerdew2014} or \citet{VanSchaeybroeckVannitsem2013}, to name just a few. It has been used as a reference technique in the recent papers by \citet{Wilks2014}, \citet{Feldmann&2014} and  \citet{ScheuererHamill2014}. 

\subsubsection{The Schaake shuffle}\label{schaake.shuffle}

In contrast to ECC, the data set determining the dependence pattern in the Schaake shuffle \citep{Clark&2004} is not given by the raw ensemble forecasts, but by past observations\\ $\mathbf{y}:=\{(y_{1}^{1},\ldots,y_{N}^{1}),\ldots,(y_{1}^{L^\ast},\ldots,y_{N}^{L^\ast})\}$ taken from $N$ different dates of a historical archive. That is, observations from the same $N$ dates are employed for all locations and weather variables throughout the procedure of the Schaake shuffle for a fixed forecast instance. Hence, we have $\mathbf{z}=\mathbf{y}$ in Scheme \ref{scheme.empcop}. In particular, $N$ does not need to equal the raw ensemble size $M$.\\
In the original implementation of the Schaake shuffle by \citet{Clark&2004}, the corresponding $N$ dates, from which the observations are taken, are chosen from all years in the historical record, except for the year of the forecast of interest, and lie within seven days before and after the verification date $t$, regardless of the year. A more general implementation of the Schaake shuffle may use observations from arbitrary (or randomly selected) past dates in the whole historical record. We will refer to this procedure as the Random Schaake method in the case study in Section \ref{case.study}.\\
The Schaake shuffle has been employed successfully in numerous applications \citep{Schaake&2007,Voisin&2010,Voisin&2011,Robertson&2013,Verkade&2013,VracFriederichs2014,Wilks2014}.

\subsection{The SimSchaake approach as a similarity-based implementation of the Schaake shuffle}\label{simschaake.subsection}

As we have seen, the Schaake shuffle generates a postprocessed ensemble inheriting the rank dependence structure from historical observations. However, its standard implementations fail to condition the multivariate dependence pattern on current or predicted atmospheric states. Inspired by the quote of \citet{Clark&2004} mentioned in the introductory Section \ref{intro}, we address this challenge by linking the Schaake shuffle to similarity- or analog-based ensemble methods. In such approaches, one seeks ensemble forecasts in an archive of past data that are similar to the current one. The basic idea is that the realizing states of the atmosphere corresponding to such an analog ensemble can be assumed to be similar to the state to be predicted \citep{HamillWhitaker2006}. These techniques have become popular and important, as for instance witnessed by the papers of \citet{BannayanHoogenboom2008}, \citet{Klausner&2009}, \citet{Hall&2010}, \citet{DelleMonache&2011}, \citet{MessnerMayr2011} and \citet{DelleMonache&2013}, and further by recent research \citep{Alessandrini&2015,Junk&2015,Vanvyve&2015}. In this context, the question of the choice of appropriate similarity criteria in a nearest neighbor sense arises, with the papers above providing some proposals.
\\
The following approach, which will be referred to as the SimSchaake method in what follows, combines the idea of searching for similar ensembles and the Schaake shuffle. Like the Schaake shuffle and in contrast to standard ECC, it can be applied to any ensemble, regardless of whether it consists of exchangeable or non-exchangeable members, and the size $N$ of the final postprocessed ensemble is not restricted to equal the raw ensemble size $M$. To describe the new SimSchaake approach formally in detail, let $\Lambda$ be the length of the training period required for the univariate ensemble postprocessing, $t_0$ the initialization date of the ensemble forecast, and $t$ the verification date. Let further $D$ be the number of dates in the past of $t_0$ for which ensemble forecast and observation data is available. For the feasibility of the SimSchaake approach, it is required that ensemble forecast and observation data is available for at least $\max\{N,\Lambda\}$ dates in the past of $t_0$, that is, $D \geq \max\{N,\Lambda\}$. Considering the prediction horizon to be fixed, the SimSchaake approach then proceeds according to the empirical copula-based postprocessing as in Scheme \ref{scheme.empcop}, where the data set $\mathbf{z}$ in step 1 is derived as follows.

\begin{enumerate}

\item[i.]{For a fixed margin $\ell^\ast$, let $\mathbf{x}^{\ell^\ast,\tau}:=(x_{1}^{\ell^\ast,\tau},\ldots,x_{M}^{\ell^\ast,\tau})$ denote the (possibly standardized) $M$-member raw ensemble  
forecast valid on date $\tau$. Let further
\begin{equation*}
\mathbf{x}^{\tau}:=(\mathbf{x}^{1,\tau},\mathbf{x}^{2,\tau},\ldots,\mathbf{x}^{L^\ast,\tau})=((x_{1}^{1,\tau},\ldots,x_{M}^{1,\tau}),(x_{1}^{2,\tau},\ldots,x_{M}^{2,\tau}),\ldots,(x_{1}^{L^\ast,\tau},\ldots,x_{M}^{L^\ast,\tau}))
\end{equation*}
denote the corresponding $(L^\ast \times M)$-tuple consisting of the $M$-member ensemble forecasts of all $L^\ast$ margins, that is, combinations of weather variable and location.
\\
If weather variables with distinct units or magnitudes are involved, the components of $\mathbf{x}^{\ell^\ast,\tau}$ should be standardized for each multi-index $\ell^\ast$.
}

\item[ii.]{
For each date $t_d$ in the past of the initialization date $t_0$, where $d \in \{1,\ldots,D\}$, compute a suitable fixed similarity criterion
\begin{equation*}
\Delta^{t_d}: \mathbb{R}^{L^\ast \times M} \times \mathbb{R}^{L^\ast \times M} \rightarrow [0,\infty),\,\,\,(\mathbf{x}^{t},\mathbf{x}^{t_d}) \mapsto \Delta^{t_d}(\mathbf{x}^{t},\mathbf{x}^{t_d}),
\end{equation*}
between the actual forecast $\mathbf{x}^{t}$ valid on the verification date $t$ and the forecast $\mathbf{x}^{t_d}$ valid on the date $t_d$. The similarity criterion $\Delta^{t_d}$ is taken to be negatively oriented, that is, the lower the similarity criterion the more similar the ensemble forecasts. A similarity criterion value of exactly zero indicates that the ensemble forecasts are identical. 
}

\item[iii.]{
Choose those $N$ dates $\tau_{1},\ldots,\tau_{N} \in \{t_1,\ldots,t_D\}$ for which the data is most similar to that for the date $t$ in the sense that the corresponding values of $\Delta^{\tau_n}$ for $n \in \{1,\ldots,N\}$ are the smallest among the values of $\Delta^{t_d}$ for $d \in \{1,\ldots,D\}$.\\
Note that the information of all multi-indices $\ell^\ast$ simultaneously is employed to determine the dates $\tau_1,\ldots,\tau_{N}$.
}

\item[iv.]{For each margin $\ell^{\ast}$, let $y^{\ell^{\ast},\tau_1},\ldots,y^{\ell^{\ast},\tau_N}$ denote the corresponding $N$ historical verifying observations valid on the dates $\tau_1,\ldots,\tau_{N}$ determined in step iii. For simplicity, write $y_{n}^{\ell^\ast}:=y^{\ell^\ast,\tau_n}$ for $n \in \{1,\ldots,N\}$ and build the data vector $\mathbf{y}^{\ell^{\ast}}:=(y_{1}^{\ell^{\ast}},\ldots,y_{N}^{\ell^{\ast}})$ for the verification date $t$. The data set $\mathbf{z}$ in step 1 from Scheme \ref{scheme.empcop} is then obtained by aggregating the historical observation databases of all margins $\ell^\ast$, that is, $\mathbf{z}=(\mathbf{y}^1,\ldots,\mathbf{y}^{L^{\ast}})$. From this template $\mathbf{z}$, the empirical copula related to the SimSchaake approach is derived via \eqref{empcop}.}

\end{enumerate}

\begin{figure}[t]
\noindent \includegraphics[scale=0.48]{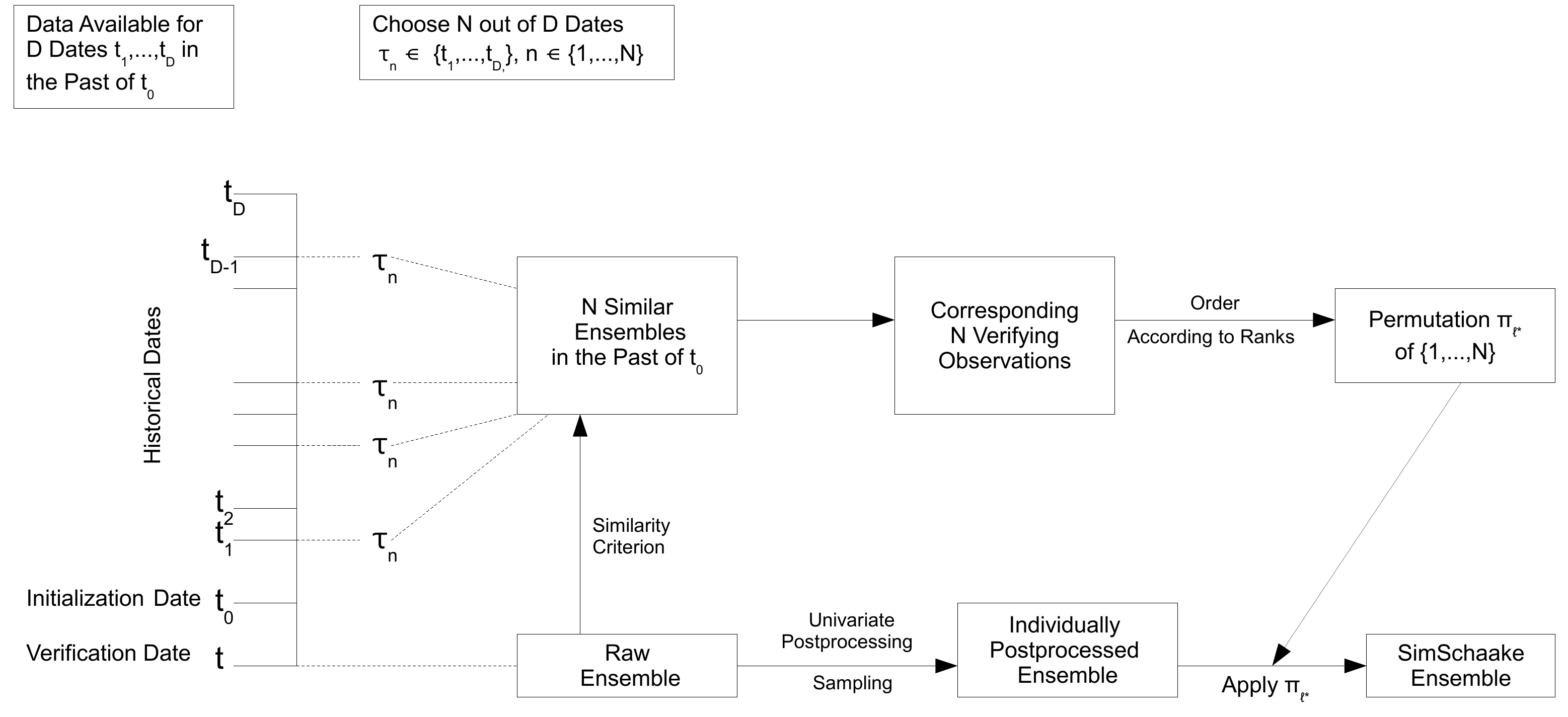}
\caption[Scheme of the SimSchaake approach]{Scheme of the SimSchaake approach}
\label{ss.scheme}
\end{figure}

\noindent  Having obtained the data set $\mathbf{z}$ and the respective empirical copula according to the above procedure, steps 2 to 4 from Scheme \ref{scheme.empcop} are performed in order to generate the final postprocessed SimSchaake ensemble. A scheme of the SimSchaake approach is given in Fig. \ref{ss.scheme}.
\\
An appropriate choice of the similarity criterion $\Delta^{t_d}$ in step ii, which is then consistently used throughout the whole SimSchaake approach, is crucial. We here consider
\begin{equation}\label{simsch.ens}
\Delta^{t_d}(\mathbf{x}^{t},\mathbf{x}^{t_d}):=\sqrt{\frac{1}{L^\ast}\sum\limits_{\ell^\ast=1}^{L^\ast}(\bar{x}^{\ell^\ast,t}-\bar{x}^{\ell^\ast,t_d})^{2}+\frac{1}{L^\ast}\sum\limits_{\ell^\ast=1}^{L^\ast}(s^{\ell^\ast,t}-s^{\ell^\ast,t_d})^{2}},
\end{equation}
where 
\begin{equation*}
\bar{x}^{\ell^\ast,\tau}:=\frac{1}{M}\sum\limits_{m=1}^{M}x_{m}^{\ell^\ast,\tau} \text{\,\,\,\,and\,\,\,\,} s^{\ell^\ast,\tau}:=\sqrt{\frac{1}{M-1}\sum\limits_{m=1}^{M}(x_{m}^{\ell^\ast,\tau}-\bar{x}^{\ell^\ast,\tau})^2}
\end{equation*}
denote the empirical mean and standard deviation, respectively, of the ensemble forecast $\mathbf{x}^{\ell^\ast,\tau}$ for the fixed multi-index $\ell^\ast$ at date $\tau$. As it does not depend on how the ensemble members are labeled, the similarity criterion in \eqref{simsch.ens} can be applied to ensembles consisting of exchangeable members. In particular, it is suitable for the temperature predictions from the European Centre for Medium-Range Weather Forecasts (ECMWF) used in the case study in the next section. While the use of the empirical mean to some degree accounts for seasonal aspects when considering temperature forecasts only, the empirical standard deviations reflect the uncertainties within the ensemble forecasts. Thus, these issues are addressed by \eqref{simsch.ens} when comparing two ensemble forecasts. Alternative proposals for similarity criteria, also for the case of ensembles with non-exchangeable members, can be found in the references mentioned before. 
\\
It is possible to transform the values of a similarity criterion $\Delta^{t_{d}}$ from $[0,\infty)$ to $(0,1]$ by employing the standardization $\tilde{\Delta}^{t_{d}}:=\exp(-\Delta^{t_{d}})$. With respect to $\tilde{\Delta}^{t_{d}}$, similarity values near to 1 indicate a very high similarity between $\mathbf{x}^{t}$ and $\mathbf{x}^{t_{d}}$, while similarity values near to 0 point at no similarity. Accordingly, if using $\tilde{\Delta}^{t_{d}}$, we then have to choose the dates $\tau_1,\ldots,\tau_{N}$ corresponding to the highest, and not to the lowest, values of $\tilde{\Delta}^{t_{d}}$ in step iii. 
\\
As mentioned, the SimSchaake approach addresses two shortcomings in the standard ECC method. First, it can also be applied to ensembles consisting of non-exchangeable members, as the reordering is not based on the ensemble forecasts, but on observations. Second, with the SimSchaake technique we can principally create ensembles of arbitrary size, as long as there are sufficiently many historical observations in the past. In contrast to ECC, the postprocessed ensemble is thus not restricted to have the same number of members as the raw ensemble.
\begin{figure}[p]
\noindent \includegraphics[scale=1]{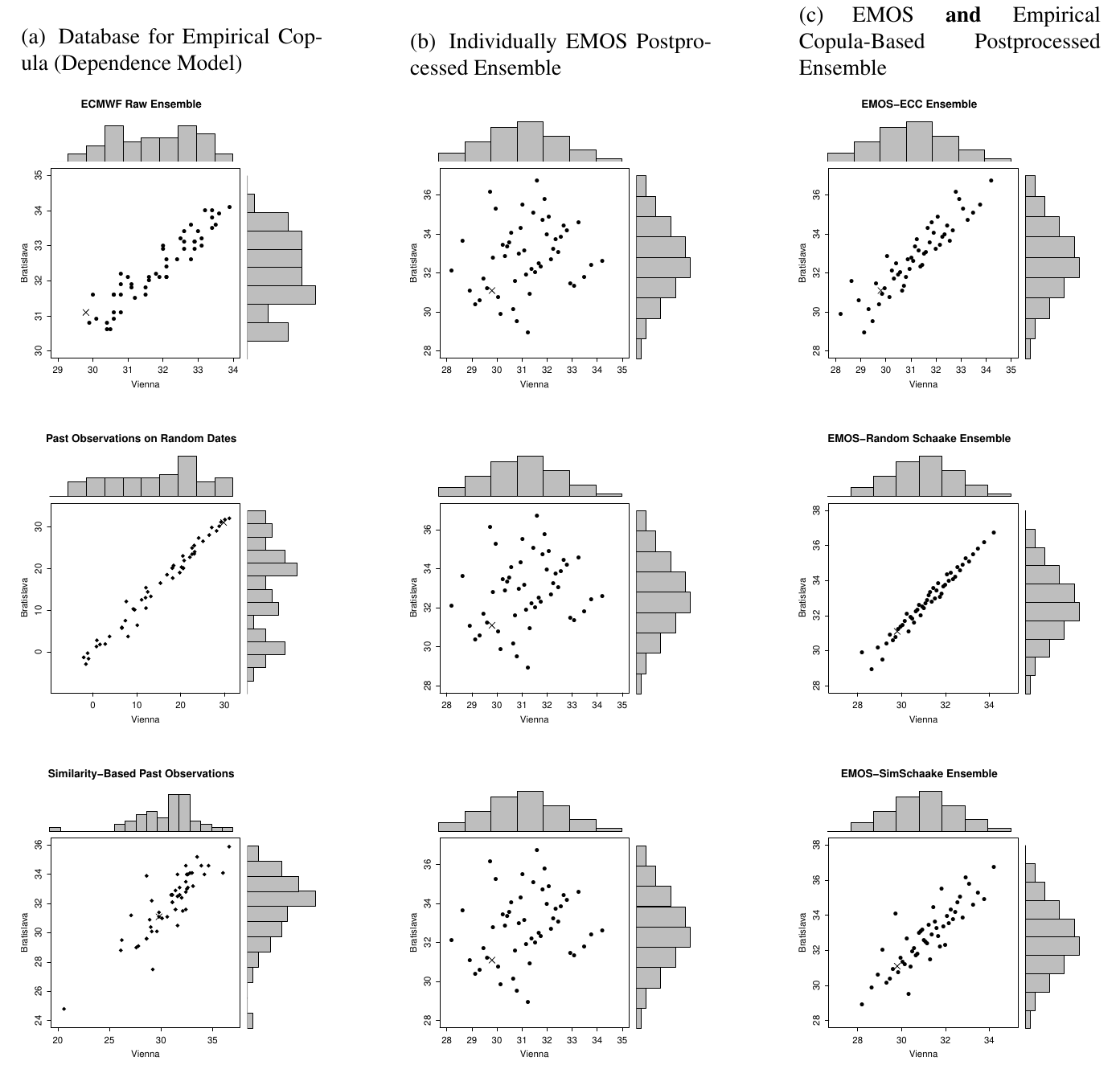}
\caption{Illustration of ECC (first row), the Random Schaake method (second row) and the SimSchaake method (third row) using 24 hour ahead temperature forecasts (in $^\circ$C) at Vienna and Bratislava valid on 9 July 2011, 1200 UTC. Ensemble forecasts are indicated by the dots, and the verifying observation by the cross. The tetragons symbolize past observations from a historical database. At the top and to the right of each subfigure, the corresponding marginal histograms are shown.}
\label{illus.empcop}
\end{figure}
\newline
\noindent In Fig. \ref{illus.empcop}, an illustration of the three empirical copula-based postprocessing methods presented in this section -- ECC in the first row, the Schaake shuffle in the second row and the SimSchaake method in the third row -- is given. We consider 24 hour ahead surface temperature forecasts (in degrees Celsius; $^\circ$C) at Vienna (Austria) and Bratislava (Slovakia) valid on the hot summer day of 9 July 2011 at 1200 UTC. In the subfigures, the ensemble forecast is indicated by the dots, the verifying observation by the cross, and past observations from a historical database by the tetragons. As the ECMWF raw ensemble used here has size $M=50$, the illustrations are for convenience based on $N=M=50$-member ensembles for both ECC and the Schaake shuffle-based methods. The first column of Fig. \ref{illus.empcop} shows the corresponding database that is used to determine the dependence structure and the empirical copula, respectively, of the postprocessing approach: the ECMWF raw ensemble in the case of ECC, past observations on random dates for the Random Schaake method, and past observations on specific dates selected according to the similarity criterion in \eqref{simsch.ens} in the case of the SimSchaake approach. The second column shows three times the same individually postprocessed ensemble forecast. This is generated by randomly pairing the equidistant quantiles \eqref{sampling.q} from the predictive CDFs obtained by univariate EMOS postprocessing \citep{Gneiting&2005} at Vienna and Bratislava separately. Such an ensemble forecast does not take account of the pronounced positive spatial correlation structure. In the third column, the final postprocessed ensemble forecasts are shown, obtained by applying the corresponding empirical copula to the samples derived from the individual EMOS postprocessing. These empirical copula-based postprocessed ensembles have the same marginal distributions as the individual EMOS postprocessed ensemble, as witnessed by the respective histograms at the top and to the right of each subfigure. Additionally, they exhibit the same spatial correlation pattern as the underlying database specified in the first column, thus respecting dependencies.

\section{Case study}\label{case.study}

\subsection{Setting}\label{data.set}

In our case study, we employ predictions of the European Centre for Medium-Range Weather Forecasts (ECMWF) core ensemble \citep{ECMWF2012}, whose $M=50$ members can be considered exchangeable. We focus on 24 hour-ahead temperature forecasts jointly at Vienna (Austria), Bratislava (Slovakia) and Budapest (Hungary), and consequently on spatial dependencies only. For each location, the ground truth is given by the corresponding surface synoptic observations (SYNOP). The approximate distance from Vienna to Bratislava is 50 kilometers, from Bratislava to Budapest 170 kilometers, and from Vienna to Budapest 210 kilometers. There are pronounced positive pairwise correlations between the observations at the three different locations, which are the stronger the closer the respective stations are. These correlation patterns are basically reflected well in the ensemble forecasts.
\\ 
We consider those 3985 test days during the period from 1 January 2003 to 31 December 2013 for which all required forecast and observation data is available at all the three stations. In our case study, all data is valid on 1200 UTC. Univariate postprocessing is performed via EMOS \citep{Gneiting&2005} using the R package {\texttt{ensembleMOS}} \citep{R2011,Yuen&2013}, employing a rolling window consisting of the last $\Lambda=50$ days before the verification day as training period. For each marginal EMOS postprocessed predictive CDF $F_\ell$, we follow the quantization \eqref{sampling.q} and take the $N$ equidistant $(n/(N+1))$-quantiles, where $n=1,\ldots,N$, as samples, focusing on the cases of $N=M=50$ and $N=80$, respectively, here.\\ 
For these desired ensemble sizes, we assess and compare the predictive performances of 
\begin{itemize}
\item{the ECMWF raw ensemble,}
\item{the Individual EMOS ensemble, which assumes independence,}
\item{the EMOS-ECC ensemble, which assumes a dependence structure according to that in the raw ensemble,}
\item{the EMOS-Random Schaake ensemble, which assumes a dependence structure according to randomly selected historical observation data, and}
\item{the EMOS-SimSchaake ensemble, which assumes a dependence structure according to specific historical observation data valid on dates for which the ensemble forecast resembled the current one with respect to similarity criterion \eqref{simsch.ens}.}
\end{itemize}
Obviously, results for the raw and the EMOS-ECC ensemble can only be reported for the case of $N=M=50$, whereas the other ensembles are additionally evaluated for the final ensemble size of $N=80$. For the two approaches employing the Schaake shuffle notion, the past dates from which the corresponding verifying observations are taken, are searched for among all available historical data, where ensemble forecast and observation data is available from 1 January 2002 to 31 December 2013. Hence, the database used for the Schaake shuffle-based methods grows over time. Recall that the EMOS-Random Schaake method just randomly selects those past dates, whereas the EMOS-SimSchaake approach chooses them based on the ensemble similarity criterion \eqref{simsch.ens}.

\subsection{Evaluation tools}\label{app.verif}
A probabilistic forecast distribution or an ensemble forecast, respectively, should be as sharp as possible, subject to calibration, which refers to statistical consistency between the forecasts and the observation \citep{Gneiting&2007}. To assess the predictive performances of our different ensembles, several verification tools are available \citep{Wilks2011}.\\
In univariate settings, calibration can be checked via the verification rank histogram \citep{Anderson1996,Talagrand&1997,Hamill2001}. As we focus on the evaluation of multivariate quantities in this paper, ensemble calibration in our case study is checked via the multivariate \citep{Gneiting&2008}, band depth and average rank histograms \citep{Thorarinsdottir&2013}. When an ensemble forecast is calibrated, the multivariate, band depth or average rank, respectively, is uniformly distributed. Calibration can thus be diagnosed by compositing over forecast cases, plotting the corresponding multivariate, band depth or average rank histogram, respectively, and checking for deviations from uniformity, that is, flatness of the histogram. For an interpretation of the different shapes a rank histogram for multivariate quantities can exhibit, see \citet{Gneiting&2008} and \citet{Thorarinsdottir&2013}.
\\
The overall forecast skill can be assessed via proper scoring rules \citep{GneitingRaftery2007}, which are able to assess calibration and sharpness simultaneously and are taken to be negatively oriented here, that is, the lower the score the better the predictive performance. A widely used proper scoring rule for univariate quantities is the continuous ranked probability score (CRPS) \citep{GneitingRaftery2007,MathesonWinkler1976}.\\
In this paper, we employ the energy score (ES) \citep{Gneiting&2008}, which is the analog of the CRPS for multivariate quantities. For an $N$-member ensemble forecast 
\begin{equation*}
\mathbf{x}_1:=(x_1^1,\ldots,x_1^L),\ldots,\mathbf{x}_N:=(x_N^1,\ldots,x_N^L) \in \mathbb{R}^L
\end{equation*}
and an observation 
\begin{equation*}
\mathbf{y}:=(y_1,\ldots,y_L) \in \mathbb{R}^L,
\end{equation*}
the ES is computed as
\begin{equation*}
\operatorname{ES}=\frac{1}{N}\sum\limits_{n=1}^{N}||\mathbf{x}_n-\mathbf{y}||-\frac{1}{2 N^2}\sum\limits_{\nu=1}^{N}\sum\limits_{n=1}^{N}||\mathbf{x}_{\nu}-\mathbf{x}_n||,
\end{equation*} 
with $||\cdot||$ denoting the Euclidean norm. As the ES reveals weaknesses in detecting misspecifications in the correlation structure \citep{PinsonTastu2013,ScheuererHamill2014}, we additionally consider the variogram score (VS) \citep{ScheuererHamill2014} to address this, which is given by
\begin{equation*}
\operatorname{VS}=\sum\limits_{\ell=1}^{L}\sum\limits_{\lambda=1}^{L} w_{\ell \lambda} \left(\big \vert y_{\ell}-y_{\lambda}\big \vert^{\frac{1}{2}}-\frac{1}{N}\sum\limits_{n=1}^{N}\big\vert x_{n}^{\ell}-x_{n}^{\lambda} \big \vert ^{\frac{1}{2}}\right)^2,
\end{equation*}
where the $w_{\ell \lambda}$'s are (optional) non-negative weights. For the case study in this paper, in which we focus on spatial dependencies, we follow the suggestion of \citet{ScheuererHamill2014} and let the weights be proportional to the inverse spatial distances between the corresponding locations. That is, we choose
\begin{equation*}
w_{\ell \lambda}:=\frac{\frac{1}{\operatorname{dist}(\ell,\lambda)}}{\sum\limits_{\genfrac{}{}{0 pt}{1}{\ell,\lambda=1}{\ell \neq \lambda}}^{L}\frac{1}{\operatorname{dist}(\ell,\lambda)}},
\end{equation*}
for $\ell \neq \lambda$ and $w_{\ell \lambda}:=0$ for $\ell=\lambda$, with $\operatorname{dist}(\ell,\lambda)$ denoting the spatial distance between location $\ell$ and location $\lambda$, where all distances have to be measured in the same unit. For the specific implementation in our $L=3$-dimensional setting here, we employ the distances between Vienna, Bratislava and Budapest as mentioned in the preceding subsection.
\\
In our case study, average scores over all forecast cases within our specific test period are reported.

\subsection{Results}\label{results}

\begin{figure}[p]
\noindent \includegraphics[scale=0.49]{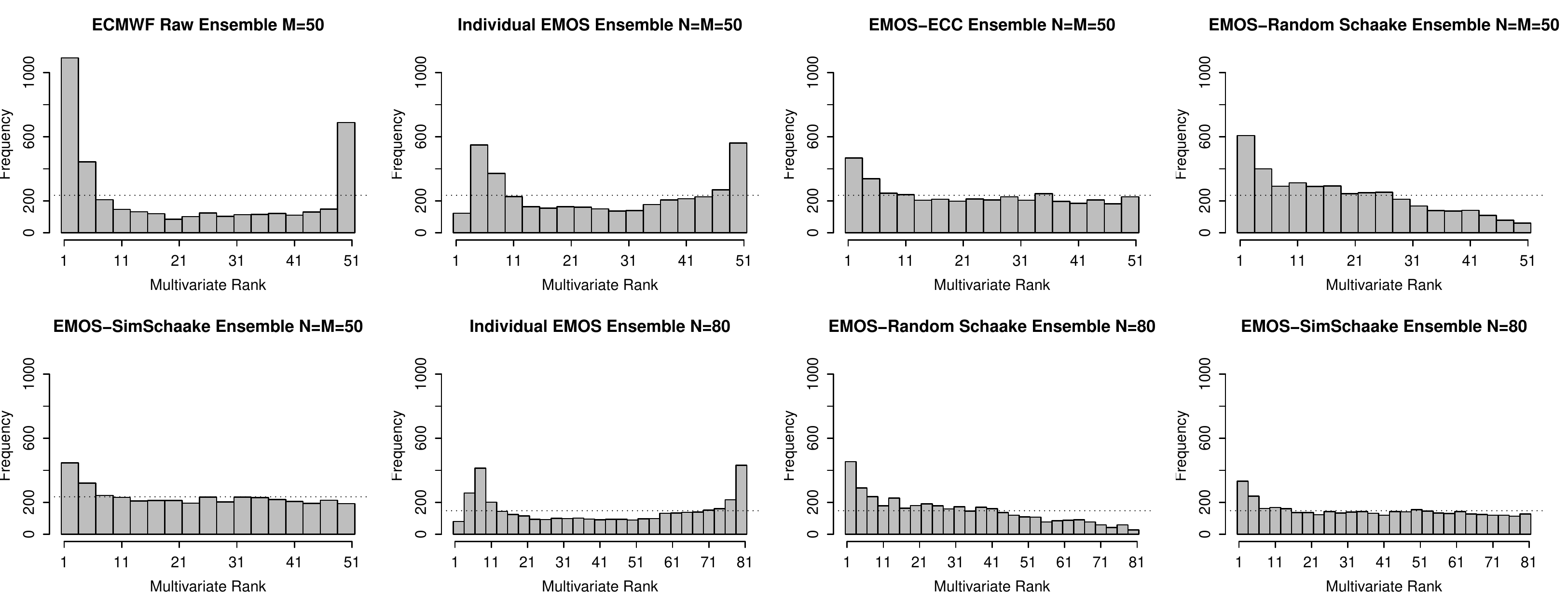}
\caption{Multivariate rank histograms for 24 hour ahead temperature forecasts at Vienna, Bratislava and Budapest jointly over 3985 test days during the period from 1 January 2003 to 31 December 2013.}
\label{mrh24}
\end{figure}

\begin{figure}[p]
\noindent \includegraphics[scale=0.49]{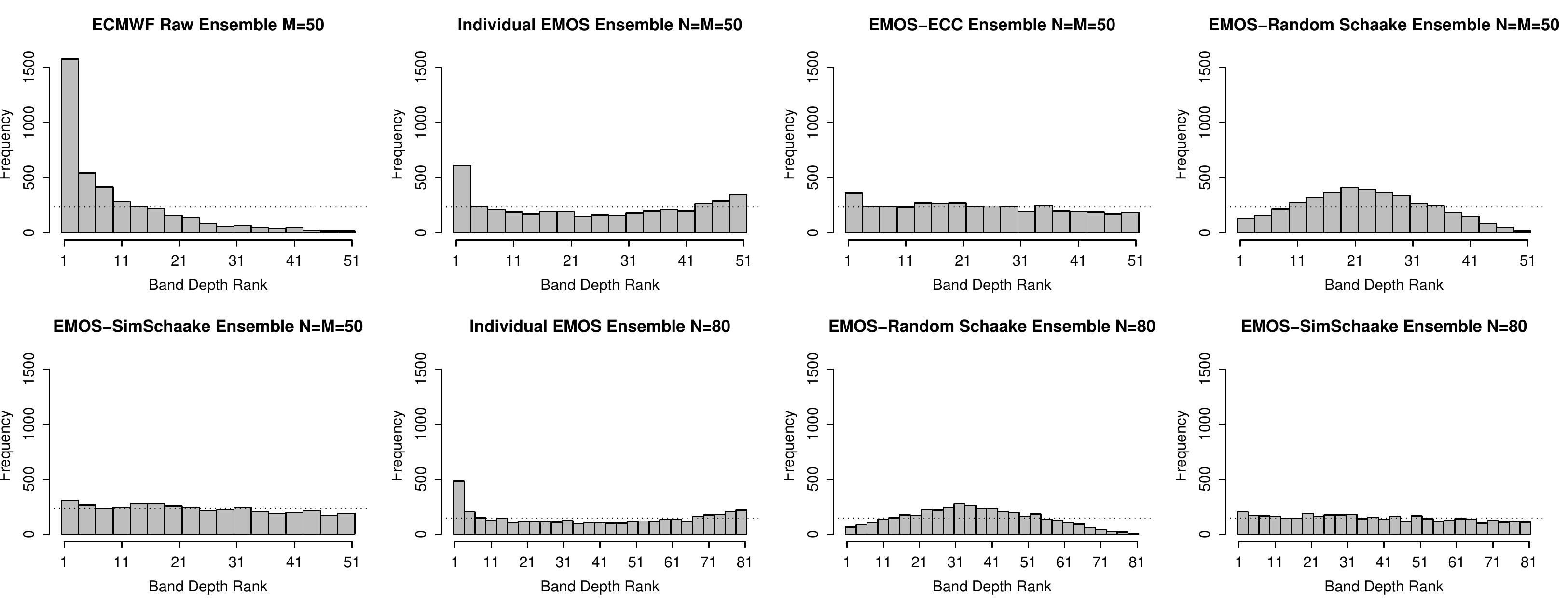}
\caption{Band depth rank histograms for 24 hour ahead temperature forecasts at Vienna, Bratislava and Budapest jointly over 3985 test days during the period from 1 January 2003 to 31 December 2013}
\label{mbd24}
\end{figure}

\begin{figure}[t]
\noindent \includegraphics[scale=0.49]{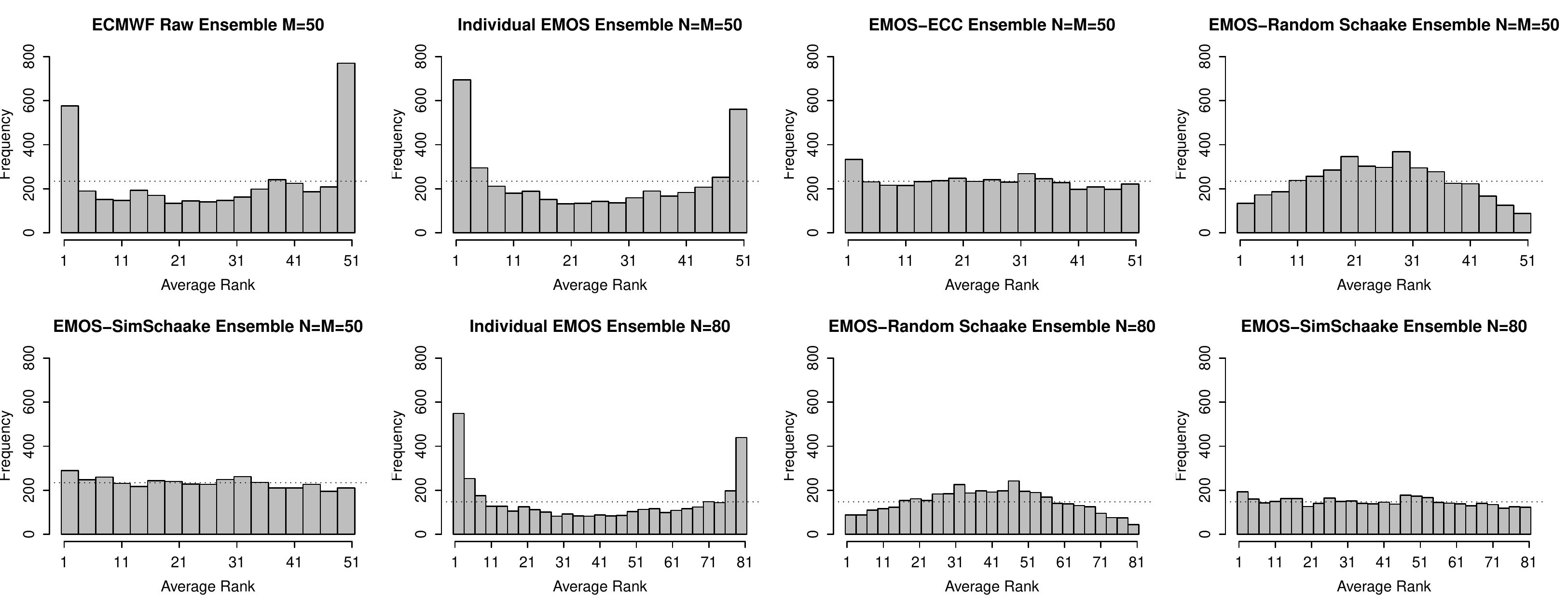}
\caption{Average rank histograms for 24 hour ahead temperature forecasts at Vienna, Bratislava and Budapest jointly over 3985 test days during the period from 1 January 2003 to 31 December 2013}
\label{avr24}
\end{figure}

\begin{table}[t]
\caption{Average energy scores (ES) and variogram scores (VS) for 24 hour ahead temperature forecasts at Vienna, Bratislava and Budapest jointly over 3985 test days during the period from 1 January 2003 to 31 December 2013. The scores for the Individual EMOS and the EMOS-Random Schaake ensembles are averaged over 100 runs.}\label{tab1}
\begin{center}
\begin{tabular}{cccc}
\hline \hline
&&ES&VS\T\B\\
\hline
$M=50$&ECMWF Raw Ensemble&2.241&0.333\T\B\\
\hline
&Individual EMOS Ensemble&1.976&0.323\T\B\\
$N=M=50$&EMOS-ECC Ensemble&1.957&0.270\T\\
&EMOS-Random Schaake Ensemble&1.998&0.300\\
&EMOS-SimSchaake Ensemble&1.952&0.265\B\\
\hline
&Individual EMOS Ensemble&1.971&0.327\T\B\\
$N=80$&EMOS-Random Schaake Ensemble&1.996&0.300\T\\
&EMOS-SimSchaake Ensemble&1.947&0.266\B\\
\hline
\end{tabular}
\end{center}
\end{table}
As we focus on the multivariate setting in this paper, we do not explicitly show the results for the univariate EMOS postprocessing at the three stations individually here. In a nutshell, the EMOS postprocessed ensemble forecasts exhibit a better predictive performance than the unprocessed raw ensemble predictions, in that they are better calibrated and have smaller CRPS values.
\\
In Table \ref{tab1}, the average energy scores (ES) and variogram scores (VS) as overall performance measures are shown. The results for the Individual EMOS and the EMOS-Random Schaake ensemble are averaged over 100 runs for each forecast instance, in order to account for randomizations. Precisely, for the Individual EMOS ensemble, the results for 100 different aggregations (that is, assignments of the member indices) of the equidistant quantiles obtained by the univariate EMOS postprocessing are averaged. In case of the EMOS-Random Schaake ensemble, the average is taken over 100 different selections of the random historical dates that are used to define the observation-based dependence model. Calibration is assessed via the multivariate, band depth and average rank histograms, respectively, in Figs. \ref{mrh24}, \ref{mbd24} and \ref{avr24}.
\newline
\noindent Both in the case of $N=M=50$ and $N=80$, all postprocessing methods outperform the raw ensemble in terms of the scores. The raw ensemble is clearly uncalibrated, more precisely underdispersed, as witnessed by the U-shaped multivariate and average rank histograms and the skewed band depth rank histogram with an overpopulation of the lowest ranks \citep{Gneiting&2008,Thorarinsdottir&2013}. 
\\
Considering the results for the postprocessed ensembles consisting of $N=M=50$ members first, the Individual EMOS ensemble is uncalibrated, yielding U-shaped rank histograms. In the case of the band depth and average rank histograms, respectively, this points at an underestimation of the correlation structure \citep{Thorarinsdottir&2013}, which is plausible, as this approach does not account for dependencies. With respect to the VS, the Individual EMOS ensemble performs worse than the other postprocessed ensembles, which assume specific correlation structures. In terms of the ES, the EMOS-ECC and the EMOS-SimSchaake ensembles outperform the Individual EMOS ensemble, while the distinctions are less pronounced than for the VS. This may be due to the discrimination inability of the ES with respect to correlations between different locations \citep{PinsonTastu2013,ScheuererHamill2014}, as hinted at in the preceding subsection.
\\
Comparing the three empirical copula-based postprocessing methods taking account of dependence patterns, the EMOS-ECC ensemble is well calibrated, apart from a slight overpopulation of the lowest ranks in all rank histograms. In contrast, the calibration of the EMOS-Random Schaake ensemble is not that good, as witnessed by the inverse U-shaped band depth and average rank histograms, indicating an overestimation of the correlation structures \citep{Thorarinsdottir&2013}. Moreover, the multivariate rank histogram of the EMOS-Random Schaake ensemble is skewed with too many low ranks. The EMOS-SimSchaake ensemble is calibrated best, with band depth and average rank histograms being close to uniform and an essentially flat multivariate rank histogram with a slight overpopulation of the lowest ranks only. The ranking of the three ensembles allowing for dependencies in terms of calibration is also reflected in the scores. Both for the ES and the VS, the EMOS-SimSchaake ensemble performs best, followed by the EMOS-ECC ensemble and finally the EMOS-Random Schaake ensemble. 
\\
The results and conclusions on calibration described above continue to hold analogously for the extended Individual EMOS, EMOS-Random Schaake and EMOS-SimSchaake ensembles comprising $N=80$ members. Similarly, the ranking of the predictive skill of these three extended postprocessed ensembles remains unchanged with respect to the ES and VS, respectively, compared to the case of $N=M=50$, with the EMOS-SimSchaake ensemble still performing by far best. The ES and VS values of the $N=80$-member ensembles are very similar to those of their counterparts in the case of $N=M=50$. Although an extension of the ensemble size is generally useful, there is no pronounced need in our case here to increase the ensemble size of $N=M=50$, which appears to be already reasonably large. 
\\
In a nutshell, the EMOS-SimSchaake ensemble based on the new method introduced in this paper performs best among all reference ensembles, both with respect to calibration and in terms of scores. In particular, the EMOS-SimSchaake ensemble outperforms the EMOS-Random Schaake ensemble. Hence, there appears to be a clear benefit of using the specific past dates on which the ensemble forecasts resembled the current one to create the historical observation database modeling the dependencies, rather than picking these dates randomly. The EMOS-SimSchaake ensemble also outperforms the EMOS-ECC ensemble.

\section{Discussion}\label{discussion}

We have discussed and compared empirical copula-based ensemble postprocessing methods that are able to account for dependencies. While ECC and the Schaake shuffle have been reviewed in a general frame, the SimSchaake scheme has been newly developed in this paper as a multivariate postprocessing tool. Essentially, the SimSchaake procedure aggregates samples from univariate postprocessed distributions, where the underlying dependence structure and the involved empirical copula, respectively, are derived from historical observations at dates in the past which showed a similar ensemble forecast to the current one. In our case study, the SimSchaake ensemble has performed best overall and better than the ECC ensemble, while having the benefit of a broader applicability, in that it can also be employed on ensembles comprising non-exchangeable members and is not restricted to have the same size as the raw ensemble.\\
The SimSchaake method depends on the design of a suitable similarity criterion, where the choice of \eqref{simsch.ens} has proven to be useful, yielding good results. However, the predictive performance of the SimSchaake ensemble might be improved by using a more sophisticatedly designed similarity criterion, perhaps including a suitable weighting function (or monotone transformations thereof) that accounts for seasonal aspects. Moreover, the similarity criterion in \eqref{simsch.ens} is tailored to ensembles consisting of exchangeable members such as the ECMWF ensemble in our case study. For ensembles comprising non-exchangeable members, other criteria might provide more reasonable and better choices.
\\
As mentioned, a drawback of the standard ECC postprocessed ensemble employed here is that it is constrained to have the same size as the unprocessed ensemble, while the Schaake shuffle and the SimSchaake ensembles are not. However, there have been first attempts to design ECC-like ensembles having more members than the raw ensemble \citep{Wilks2014,Schefzik2015}. Similar to the work in \citet{Wilks2014}, an issue for future work may be to design and conduct a further case study including these approaches to allow for a broader comparison of ECC- and Schaake shuffle-based concepts.

%
\section*{Acknowledgments}
I gratefully acknowledge support by VolkswagenStiftung under the project 
``Mesoscale Weather Extremes: Theory, Spatial Modeling and Prediction''. Initial work on this paper was done during my time as Ph.D. student at Heidelberg University, funded by Deutsche Forschungsgemeinschaft through Research Training Group (RTG) 1953. I thank Tilmann Gneiting, Stephan Hemri, Sebastian Lerch and Martin Leutbecher for valuable comments, suggestions and discussions. The forecast data used in the case study have been made available by the European Centre for Medium-Range Weather Forecasts. I thank Stephan Hemri for help with the data and Michael Scheuerer for R code for the rank histograms for multivariate quantities.

%






%
%
%

\bibliographystyle{plainnat}
\bibliography{biblio_simschaake}

\end{document}